\documentclass[prb,twocolumn,showpacs,aps]{revtex4}
\usepackage{graphicx}
\begin{document}

\preprint{cond-mat/0000000}

\title{Study of the electron-phonon interaction in metal diborides
MeB$_{2}$ (Me=Zr, Nb, Ta, Mg) by point-contact spectroscopy}
\vspace{1cm}
\author{I. K. Yanson$^1$, Yu. G. Naidyuk$^1$,
O. E. Kvitnitskaya$^1$, V. V. Fisun$^1$, N. L. Bobrov$^1$, P. N.
Chubov$^1$, V. V. Ryabovol$^1$, G. Behr$^2$, W. N. Kang$^3$, E.-M.
Choi$^3$, H.-J. Kim$^3$, S.-I. Lee$^3$, T. Aizawa$^4$, S.
Otani$^4$, and S.-L. Drechsler$^2$}

\affiliation{$^1$ B. Verkin Institute for Low Temperature Physics
and Engineering, National Academy  of Sciences of Ukraine,  47
Lenin Ave., 61103,  Kharkiv, Ukraine}
 \affiliation{$^2$ Leibniz-Institut f\"ur Festk\"orper- und Werkstofforschung
Dresden e.V., Postfach 270116, D-01171 Dresden, Germany}
 \affiliation{$^3$ Pohang University of Science and Technology,
Pohang 790-784, South Korea}
 \affiliation{$^4$ Advanced Materials Lab., National Institute
for Materials Sciences, 1-1 Namiki, Tsukuba, Ibaraki 305-0044,
Japan}

\date{\today}

\begin{abstract}
We review investigations of the electron-phonon interaction (EPI)
in metal diborides MeB$_{2}$ (Me=Zr, Nb, Ta, Mg) by point-contact
(PC) spectroscopy. For transition metal compounds the PC EPI
functions were recovered and EPI parameter $\lambda\lesssim 0.1$
were estimated. The data are consistent with the measured surface
phonon dispersion curves. The low $\lambda$ value questions some
reports about superconductivity in these compounds. Contrary, EPI
in superconducting MgB$_2$ films manifests also in the PC spectra
itself by virtue of an elastic EPI contribution to the excess
current determined by the energy dependence of the superconducting
order parameter. To analyse the phonon features in the PC spectra
of MgB$_2$ a two-band model is exploited and the proximity effect
in the {\bf k}-space is suggested.

\end{abstract}

\maketitle

{\it Introduction}. Recently discovered superconductivity in the
{\it sp} compound MgB$_{2}$\cite{Akimitsu} attracts much attention
mainly due to  relatively high $T_{c}\simeq 39$\,K, which is the
highest as for two-component systems. Probably a specific feature
of this compound appears to be a rare example of two disconnected
bands of the Fermi surface with quite different dynamical
properties\cite{Shulga,Mazin,Liu}. One of those bands is
two-dimensional (2D), with extremely strong electron-phonon
interaction (EPI), while the other is 3D with a weak EPI \cite
{Mazin,Liu,Kong,Yildirim}.

The overwhelming majority of the community anticipated an
electron-phonon mechanism in MgB$_{2}$
\cite{Liu,Kong,Yildirim,Choi,Bohnen}. Thus, the preferable reason
of the formation of Cooper pairs is EPI, and its investigation to
understand the nature and peculiarities of superconductivity in
MgB$_{2}$ is vastly desirable.

 In the wake of the finding of superconductivity in MgB$_2$
scientific activity was applied to its search in other diborides
(all having a hexagonal layered crystal structure of an AlB$_2$
type) along with the further study of their properties. According
to a recent review\cite{Buzea} no superconducting transition has
been observed so far in the diborides of transition metals MeB$_2$
(Me=Ti, Zr, Hf, V, Cr, Mo). As to the superconductivity in
NbB$_2$, TaB$_2$ and ZrB$_2$ controversial reports can be
found\cite{Buzea,Gaspa}.

In this article we would like to summarize our findings in the
determination of the EPI function for the above mentioned
diborides from the point-contact (PC) spectra. The attractive
feature of the PC spectroscopy is that the measurement at low
temperatures of the voltage derivative of the PC resistance allows
a straightforward determination\cite{Yanson} of the PC EPI
function $\alpha^2F(\omega)$. The knowledge of $\alpha^2F(\omega)$
for conducting systems provides a consistent check for the
possibility of a phonon-mediated pairing mechanism, e.g., by
estimation of the electron-phonon-coupling strength characterized
by the EPI parameter $\lambda = 2 \int \alpha^2
F(\omega)\omega^{-1}d\omega$. From the comparison of the
experimentally determined $\alpha^2F(\omega)$ with theoretical
calculations different models and approaches can be discriminated.

{\it Experimental details}. The samples are  {\it c}-axis oriented
films of MgB$_{2}$, whose characteristics are described in
Ref.\cite{Kang} and single crystals of MeB$_{2}$
(Me=Zr,\,Nb,\,Ta). The residual resistivity $\rho_0$ and the
residual resistivity ratio (RRR) of all investigated diborides are
shown in Table 1.

Different PC's  were established {\it in  situ} by touching  of
the MgB$_2$ film or the cleaved surface of a transition metal
single crystal by a sharpened edge of an Ag counterelectrode. This
geometry corresponds to a current flowing preferably along the
c-axis of the film. A number of contacts was measured by touching
the film edge after breaking Al$_2$O$_3$ substrate. By this means,
the current flows preferably along the ab plane. As to PC's on
another diborides, the orientation of electrodes was not
controlled with respect to the crystallographic axis. The
zero-bias resistance $R_0$ of the investigated contacts ranged
from a few ohms up to several tens of ohms at 4.2\,K.

Both the differential resistance d$V/$d$I$ and the second
derivative of $I-V$ characteristic d$^2V$/d$I^2(V)$ vs $V$ were
registered using a standard lock-in technique.

\begin{table}\label{yanst1}
\caption{Parameters of investigated  diborides.}
\begin{tabular} {cc|ccccccccc} \hline 
Compound & & & Sample & & $\rho_0\,\cdot10^9$, $\Omega$\,m & & RRR
& &
 $n ^a$\,$\cdot 10^{-28}$, m$^{-3}$& \\ \hline
 MgB$_2$& & & film & & 60 & & 2.3 & & 6.9 & \\
 ZrB$_2$& & & sgl. cry. & &  3.3 & & 24 & & 13 & \\
 NbB$_2$& & & sgl. cry. & &  - & & - & & 18.2 & \\
 TaB$_2$& & & sgl. cry. & &  220 & & 1.2& & 18.6 & \\ \hline 
\end{tabular}
$^a$ The density of carriers $n$ was estimated by the number of
valence electrons (2 for MgB$_2$, 4 for ZrB$_2$, 5 for NbB$_2$ and
TaB$_2$) per volume of the corresponding unit cell.
\end{table}

{\it Results and discussion}. According to the Kulik, Omelyanchouk
and Shekhter theory\cite{KOS} the second derivative
-\,d$^2I$/d$V^2(V)$ of the $I-V$ curve of a ballistic PC at low
temperatures due to EPI is proportional to $\alpha_{\rm
PC}^2\,F(\omega)$. In the free electron approximation\cite{Kulik}
\begin{equation}
\label{pcs} -\frac{{\rm d}^2I}{{\rm d}V^2}\propto R^{\rm
-1}\frac{{\rm d}R}{{\rm d}V}= \frac{8\,{\rm e}d}{3\,\hbar v_{\rm
F}}\alpha_{\rm PC}^2(\epsilon)\,F(\epsilon)|_{\epsilon={\rm e}V} ,
\end{equation}
where $R={\rm d}V/{\rm d}I$. The factor
$K=1/2(1-\theta/\tan\theta)$, entering in $\alpha_{\rm PC}^2$,
takes into account the kinematic restriction of electron
scattering processes in PC (for transport and Eliashberg EPI
functions the corresponding factors are: $K=(1-\cos\theta)$ and
$K$=1, respectively, with $\theta$ is the angle between initial
and final momenta of scattered electrons). Respectively, the large
angle $\theta \to \pi$ scattering (back-scattering) processes of
electrons dominate in $\alpha_{\rm PC}^2$.

According to Eq.(\ref{pcs}) the measured rms signal of the first
$V_1$ and second $V_2$ harmonics of a small alternating voltage
superimposed on the ramped $dc$ voltage $V$ defines the EPI
function $\alpha_{\rm PC}^2(\epsilon)\,F(\epsilon)$:
\begin{equation}
\label{pcs1} \alpha_{PC}^2(\epsilon)\,F(\epsilon)=
\frac{3\sqrt{2}}{4}\frac{\hbar v_{\rm F}}{{\rm
e}d}\frac{V_2}{V_1^2} .
\end{equation}
To estimate the PC diameter $d$, which enters in Eqs.(\ref{pcs})
and (\ref{pcs1}), the relation $R_{\rm PC}(T) \simeq  \frac {16
\rho l}{3\pi d^2} + \frac{\rho (T)}{d}$ derived by
Wexler\cite{Wexler}  is commonly used, which consists of a sum of
the ballistic, Sharvin, and the diffusive, Maxwell, terms. Here
$\rho l = p_{\rm F}/n$e$^2$, where $p_{\rm F}$ is the Fermi
momentum, $n$ is the density of charge carriers.

\begin{figure}
\begin{center}
\includegraphics[width=9cm,angle=0]{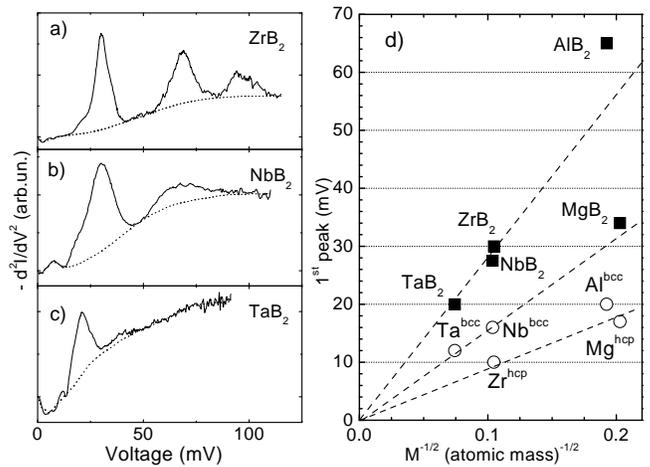}
\end{center}
\caption[] {Left panel: PC spectra -\,d$^2I(V)$/d$V^2\propto$
d$^2V$/d$I^2$(d$V/$d$I$)$^{-3}$ for investigated compounds at
$T=4.2$K. The dashed curves represent background behavior. The
zero-bias resistance and modulation signal are for ZrB$_2$:
$R_0=5.5\,\Omega$, $V_1(0)$=0.8\,mV; for NbB$_2$:
$R_0=50\,\Omega$, $V_1(0)$=2.8\,mV; for TaB$_2$: $R_0=25\,\Omega$,
$V_1(0)$=1.3\,mV. Right panel: The position of the first peak
(squares) in the measured spectra for ZrB$_2$, NbB$_2$ and TaB$_2$
vs the inverse square root of atomic mass of the transition
metals. For MgB$_2$ and AlB$_2$ the peak position is according to
the inelastic neutron scattering data\cite{Renker}. Open circles
show the position of the first peak in the PC spectra for the
corresponding metals\cite{Zr,Khot}. Straight dashed lines are
guide to eye. } \label{yansf1}
\end{figure}

\subsection{Transition metal diborides}
Representative examples of the d$^2I$/d$V^2(V)$ dependencies for
each compound averaged over both voltage polarities  are shown in
Fig.\,\ref{yansf1} (left panel). Reproducible phonon maxima are
clearly resolved up to 100\,mV (see ZrB$_2$), while for TaB$_2$
only low energy peak at 20\,mV is seen. The spectra exhibit also a
zero-bias anomaly, especially remarkable in NbB$_2$ and TaB$_2$. A
common feature for all spectra is the presence of the main
low-energy maximum placed at about 30, 28 and 20\,mV for ZrB$_2$,
NbB$_2$ and TaB$_2$, respectively. This is in line with the
general consideration that at fixed spring constants the phonon
frequency decreases with increasing of atomic mass (see
Fig.\,\ref{yansf1}, right panel). Such a behavior suggests that
the first peak corresponds to the vibration of the transition
metal.  On the other hand the neutron data peak position for
MgB$_2$ on Fig.\,\ref{yansf1} (right panel) at about
35\,meV\cite{Renker} is far below the straight line connecting the
MeB$_2$ compounds. This might be considered as the consequence of
a softening of the corresponding spring constants (i. e. metallic
bonds in MgB$_2$ instead of relatively strong Me$d$B$2p$ covalent
bonds in the MeB$_2$ series).

\begin{table}[b]
\label{yanst2} \caption{The phonon maxima and the EPI constant
$\lambda$ in MeB$_2$ compounds measured by PC spectroscopy. The
fifth column shows the maximal energy for phonon features in the
PC spectrum\cite{Zr,Khot} for the corresponding transition metals:
Me=Zr, Nb, Ta.}
\begin{tabular}{c|ccccc} \hline 
 Samples & 1$^{st}$ peak &
 2$^{nd}$ peak & 3$^{d}$ peak &$\hbar\omega_{max}^{{\rm
 T}}$& $\lambda_{\rm PC}$\\
 & meV & meV &meV & meV& \\
 \hline
 MgB$_2$ & $30(?)$ &  $\sim 60$ &  $\sim 90$ & 30& -\\
 ZrB$_2$ & $30\pm 0.5$ &  $68\pm 1$ &  $94\pm 2$ & 25& 0.06\\
 NbB$_2$ & $28\pm 2$ & $60\pm 5$ & - & 28& 0.08\\
 TaB$_2$ & $20\pm 1$ & 40(?) & - & 20 & 0.025\\ \hline 
\end{tabular}
\end{table}

\begin{figure}
\begin{center}
\includegraphics[width=9cm,angle=0]{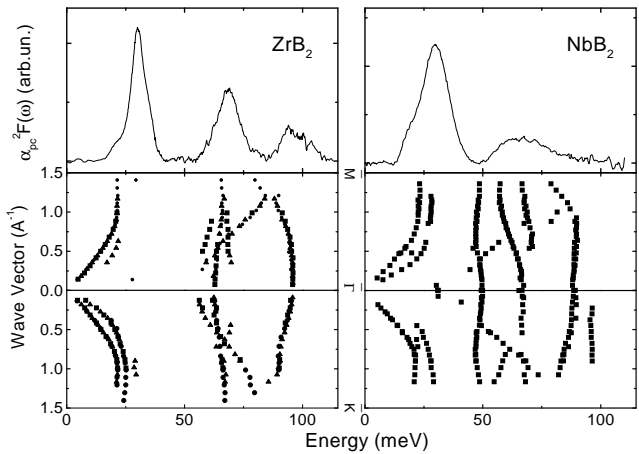}
\end{center}
\caption[] {The PC EPI function for ZrB$_2$ and NbB$_2$ recovered
from the spectra in Fig.\,\ref{yansf1}. (top panel) as compared to
the HREELS data for these compounds\cite{Aizawa} (bottom panel).}
\label{yansf2}
\end{figure}

Fig.\,\ref{yansf2} presents the recovered PC EPI functions (top
panel) for the investigated diborides ZrB$_2$ and NbB$_2$
calculated by Eq.(\ref{pcs1}) along with the High Resolution
Electron-Energy-Loss Spectroscopy (HREELS) data\cite{Aizawa} for
surface phonon dispersion (bottom panel).  Details of the
reconstruction of the PC EPI functions according to
Eq.(\ref{pcs1}) are described in Ref.\cite{NaidyukPRB}.

Because of the large mass difference between the transition metal
and the boron atoms and strong covalent B-B bonds the boron modes
are expected at much higher energy. For ZrB$_2$ two additional
maxima at 70 and 100\,mV are well resolved, while for NbB$_2$ the
high energy part of the spectrum presents a broad maximum around
60\,mV. For TaB$_2$ the high energy phonon peaks were difficult to
resolve, where according to a rough estimation\cite{Rosner} the
boron in-plane and out of plane displacement modes should have
energies of 98 and 85\,meV, respectively. No spectral features
were found for the mentioned compounds above 100\,meV. Our spectra
are in line with the measured surface phonons\cite{Aizawa} for
ZrB$_2$ and NbB$_2$ although the surface phonons, in general, are
softer. For both compounds phonon dispersion study\cite{Aizawa}
(Fig.\,\ref{yansf2}, bottom panel) demonstrates a gap between
30-50\,meV which separates acoustic and optic branches. Close to
this energy region a minimum in our PC spectra occurs. Comparing
the high energy parts of the ZrB$_2$ and NbB$_2$ PC spectra, we
may support the statement\cite{Aizawa}, that for NbB$_2$ the boron
surface phonon modes are softer and more stretched than in the
case of ZrB$_2$ giving rise to structureless maxima around
60-70\,mV (see Fig.\,\ref{yansf1}). Note that the upper boundaries
of the ZrB$_{2}$ and NbB$_{2}$ PC spectra are at about 110\,mV and
90\,mV (see Fig.\,\ref{yansf1}), what is much larger than their
Debye temperatures of 280\,K and 460\,K, respectively, estimated
from the Bloch-Gr\"uneisen temperature dependence of the
resistivity\cite{Gaspa}.

With the use of the EPI function  the parameter $\lambda = 2 \int
\alpha^2 F(\omega)\omega^{-1}d\omega$ was calculated. For many
superconductors it was found\cite{Khot} that $\lambda_{\rm PC}
\simeq \lambda_{Eliashberg}$. Table 2 shows that $\lambda_{\rm
PC}$ is rather low for the investigated diborides what is in line
with the  small $\lambda$ values reported for PC studies of the
transition metal silicides NbSi$_2$ and TaSi$_2$\cite{Balkas}
($\lambda_{\rm PC}\simeq$0.02). The possible reasons for a  small
$\lambda$ value are discussed in Ref.\cite{NaidyukPRB}. Our
results  show that NbB$_2$ has the largest $\lambda_{\rm PC}$
among the studied compounds, therefore existence of
superconductivity in NbB$_2$ with observable $T_c$ is more likely.

\subsection{MgB$_2$}
\begin{figure}
\begin{center}
\includegraphics[width=9cm,angle=0]{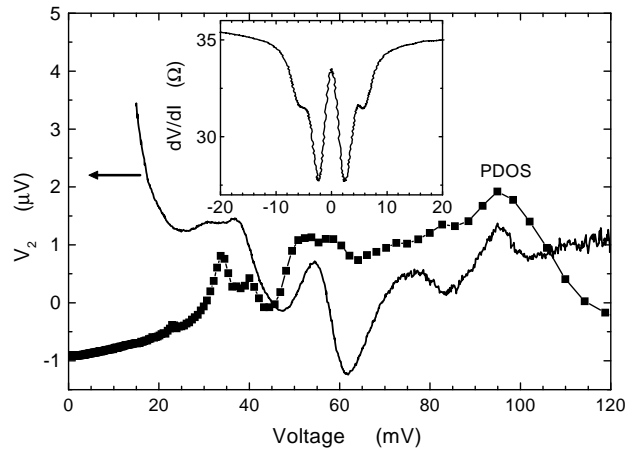}
\end{center}
\caption[] {Averaged for both polarities  spectrum $V_2 \propto
d^2V/dI^2$ after subtracting of a linear background  of MgB$_2$-Ag
PC  (R$_0=35\,\Omega$, T=4.2\,K,  V$_1$(0)=2.52\,mV) in comparison
to the neutron PDOS of MgB$_2$ \cite{Renker}. Inset: differential
resistance R=$dV$/$d$I($V$) of the same PC. } \label{yansf3}
\end{figure}

In Fig.\,\ref{yansf3} the $V_{2}\left( eV\right)\propto
d^{2}V/dI^{2}\left( eV\right)$ and $dV/dI(V)$ (in the inset)
characteristics are shown in the superconducting state at zero
field. In the energy-gap region the
two gap features are clearly seen with $\Delta _{1}=2.4$ and $%
\Delta _{2}=5.7$ meV (see inset). Further on, we simply denote by
$\Delta $ the position of a $dV/dI$ minimum. In Fig.\,\ref{yansf3}
the $V_2(V)\propto$ d$^2V$/d$I^2$ dependence after subtracting of
linear background in comparison to the phonon density of states
(PDOS)\cite{Renker} is shown. Some accordance is seen between the
positions of PDOS peaks and maxima in $V_2(V)$ dependence.

\begin{figure}
\begin{center}
\includegraphics[width=9cm,angle=0]{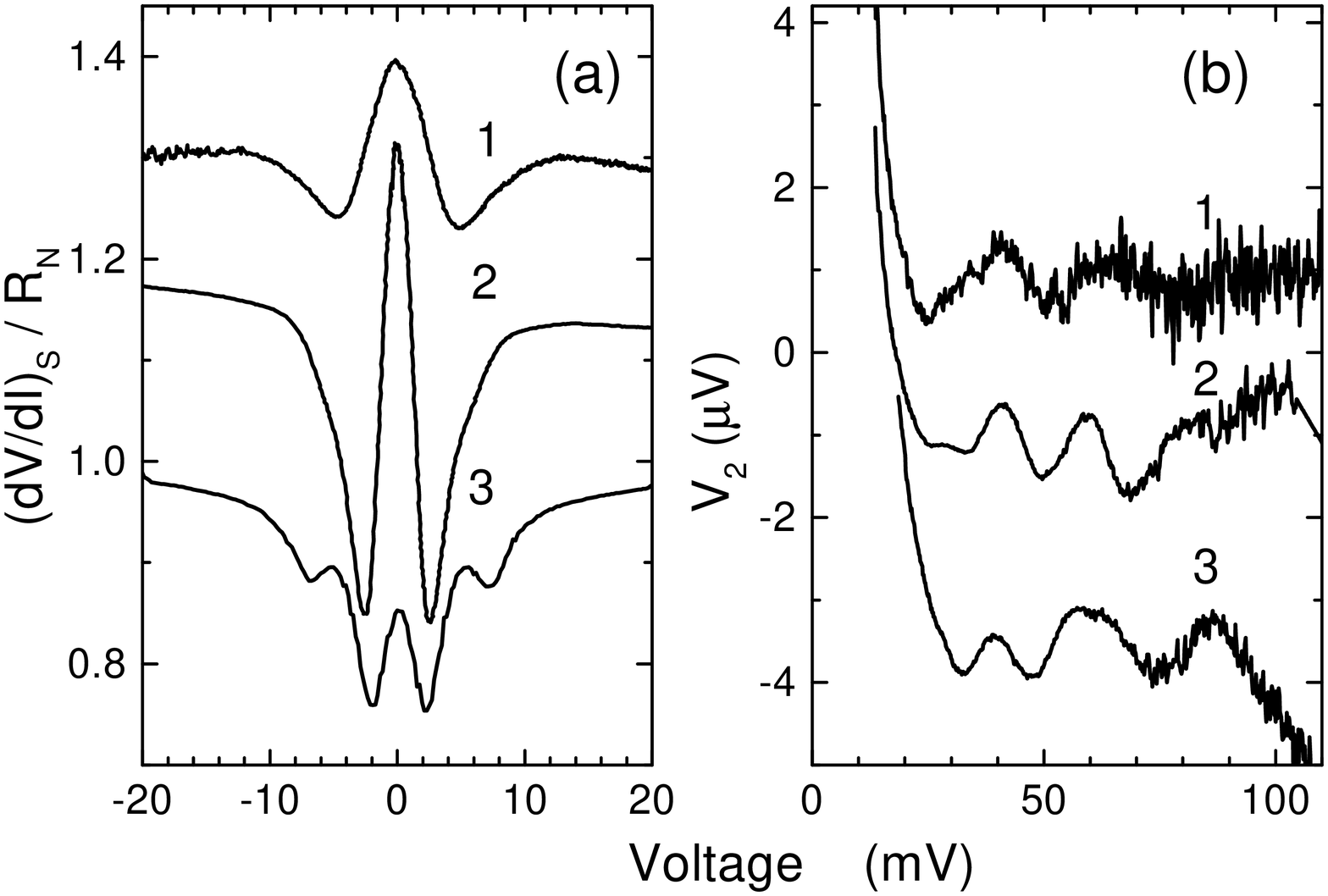}
\end{center}
\caption[] {(a) Normalized to the normal state (at $V \simeq$
30\,mV) differential resistance at $T$=4.2\,K  for 3 PC's with
$R_0$=111, 43, and 45\,$\Omega$ (curves 1-3 respectively). Curves
1,2 are shifted for clarity. (b) Second harmonic signal $V_2$ with
modulation voltages $V_1$(0)=2.5, 2.78, and 3.31\,mV, respectively
for same PC's.} \label{yansf4}
\end{figure}

The reproducibility of this kind of PC spectrum for different
contacts is seen in Fig.\,\ref{yansf4}(b). Here, the contact
resistance varies from 43 to 111 $\Omega $ with gap minima at
$\Delta =$ 2.1, 2.6, and 4.9\,meV, respectively. The larger gap
for curve 3 equals 7.0\,meV. All the PC spectra (b) correlate with
PDOS. The slight variation is probably due to the anisotropy and
different scattering rates in the contacts. Compared with the
theoretical EPI function for an isotropic one-band model (see, for
example, Ref.\cite{Kong,Yildirim,Bohnen}), the peak at eV=60$\div
$70 meV of $E_{2g}$ boron mode is not too much higher in
intensity, in accord with our earlier observation\cite{Bobrov}.
For the two-band model in clean limit, we expect an information
mostly from the 3D band keeping in mind that we measure the c-axis
oriented films. In this case, the theory predicts, that the EPI
spectral function has the $E_{2g}$-peak intensity of the same
order of magnitude as the other peaks of PDOS (see Fig.\,1 in
Ref.\cite{Golubov})

According to our classification of the energy gap
structure\cite{Naidyuk}, it is due to the random orientation of
the contact axis and scattering of charge carriers between
two-band of Fermi surface, having gaps $\Delta _{1}$ and $\Delta
_{2}$ mixed. With an increase of interband scattering the
magnitude of $\Delta_1$ ($\Delta_2$) moves to the higher (lower)
value, respectively.  For dirty contacts, where the admixture of
the 2D band is essential, only one gap maximum remains with a
broad distribution around $\simeq3.5$ meV\cite{Naidyuk}. In this
case $l$ is smaller than $d$ (where $l$ and $d$ are the electron
mean free path and the size of the contact, respectively), and the
inelastic backscattering contribution to the phonon structure,
proportional to $l/d$\cite{KulYan},  is small. Therefore, in the
superconducting state, the observed phonon structure presents
mainly the elastic contribution to the excess
current\cite{YansonSC}. The elastic term is proportional to the
energy dependent part of the excess current
$I_{exc}\left(eV\right)$, similar to the phonon structure in the
quasiparticle DOS for tunneling
spectroscopy\cite{YansonSC,Beloborod'ko}:

\begin{equation}
\frac{dI_{exc}}{dV}(eV)= \frac{1}{R_{0}}\left|\frac{\Delta
_{in}\left(
eV\right) }{eV+\sqrt{\left( eV\right)^{2} -\Delta _{in}^{2}\left( eV\right) }%
}\right|^{2} ,  \label{elastic}
\end{equation}
where $\Delta _{in}\left( eV\right) $ is the gap parameter in the
3D band, induced by the 2D-band EPI. As seen in
Fig.\,\ref{yansf4}\,(a), if the $\Delta$ value increases (curves
1-3), then the intensity
of the phonon structure in the units of $d(\ln R)/dV=2\sqrt{2}%
V_{2}/V_{1}^{2}$ increases too (Fig.\,\ref{yansf4}(b)). That is
because the modulation voltage $V_1$ decreases, whereas the
amplitudes in $V_2(eV)$-units remain approximately the same. Note,
that for curve 3 in Fig.\,\ref{yansf4}(a) one has to take the
lower gap $\Delta_1$, because the current is mainly determined by
3D band, as mentioned above.

\begin{figure}
\begin{center}
\includegraphics[width=9cm,angle=0]{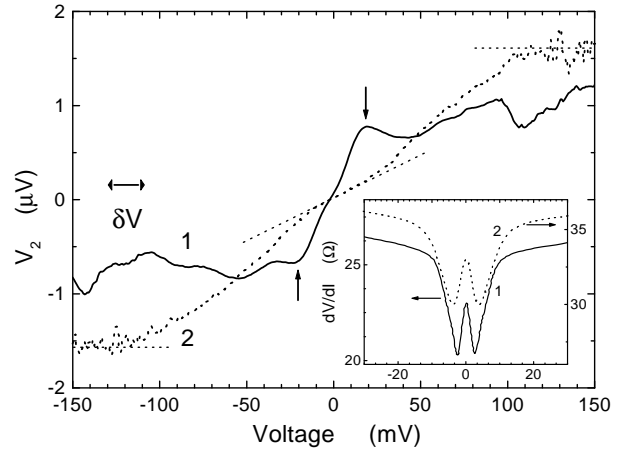}
\end{center}
\caption[] {Normal state spectra of MgB$_2$-Cu PC's taken at 41\,K
with V$_1$(0)=2.2\,mV. For spectrum No.1 vertical arrows mark the
low frequency bumps at about 20\,mV. The horizontal double arrow
bar stands for thermal smearing, determining the spectral
resolution $\delta$V.} \label{yansf5}
\end{figure}

In Fig.\,\ref{yansf5}, the normal state spectrum above $T_c$ is
displayed together with the differential resistances in the
superconducting state (see inset), showing the energy gap
structure. In spite of an increase of the temperature smearing
above 20\,mV at 41\,K the residual phonon structure is still
visible (curve 1 in Fig.\,\ref{yansf5}). The smeared phonon
features are superimposed on the rising linear background. For
curve 2, we see an increase in scattering at $\simeq 35$\,mV,
where the acoustic phonon peak occurs, and the saturation around
100\,mV, where the phonon spectrum ends. In the normal state, only
those nonlinearities remain, which are due to the inelastic
processes\cite{KOS}.  We stress that judging from the larger
superconducting gap value ($\simeq 3.5$ meV for curve 2 in the
inset) the essential contribution is expected from the $\Delta
_{2}(E)$ of 2D band for normal-state spectrum No.2. This spectrum
should contrast with curve 6 in Fig.\,\ref{yansf6} (see below),
where the direct contribution from the 2D band is small, due to
lower value of $\Delta$. Thus, the backscattering processes from
the 3D band are mostly essential for spectrum 6 and the phonon
features are not resolved at high energies. The shape of spectrum
1 in Fig.\,\ref{yansf5} presents the behavior intermediate between
these two extremes, although its energy gap is approximately equal
to curve 6 in Fig.\,\ref{yansf6}. Beside the visible phonon
features in the range $30 \div100$\,mV, it possesses a low energy
bump at about 20\,mV.

The phonon spectra of PC with a small value of energy gaps are
characterized by the presence of low frequency phonon peaks. The
small peak at energy of about 25\,mV (Fig.\,\ref{yansf6}, curves
1, 2) is visible, where a tiny knee exists on the
PDOS\cite{Renker} (see PDOS in Fig.\,\ref{yansf3}). In the normal
state (at $T\geq T_{c}$), these low frequency peaks transform into
the S-shape structure in d$^{2}V/$d$I^{2}(V)$ spectra (curve 6),
corresponding to the wide minimum of d$V/$d$I(V)$ near zero bias.
This low-frequency structure is hardly due to the remnants of
superconducting quasigap at $T>T_{c}$, since it is absent in
junction No.\,2, whose characteristic is shown in
Fig.\,\ref{yansf5}. Rather, it could be thought of as strong
interaction in the 3D band with very low-frequency excitations,
whose origin is not clear yet. On increasing the temperature, the
low frequency peak broadens further like common spectral features
do in the normal state.

Fig.\,\ref{yansf6} shows (compare curves 1 and 2) that the field
smears out the intensity of the high energy peaks, which are
induced in the 3D band by EPI of the 2D band. The disappearance of
phonon peaks at field and temperature rise proves that they do not
belong to the inelastic backscattering processes, which should
have the same intensity both in the superconducting and in normal
states. It seems more plausible that the high energy phonon peaks
are due to the elastic contribution in the excess current
(\ref{elastic}) induced by EPI from the 2D band, as was already
stated above.

\begin{figure}
\begin{center}
\includegraphics[width=9cm,angle=0]{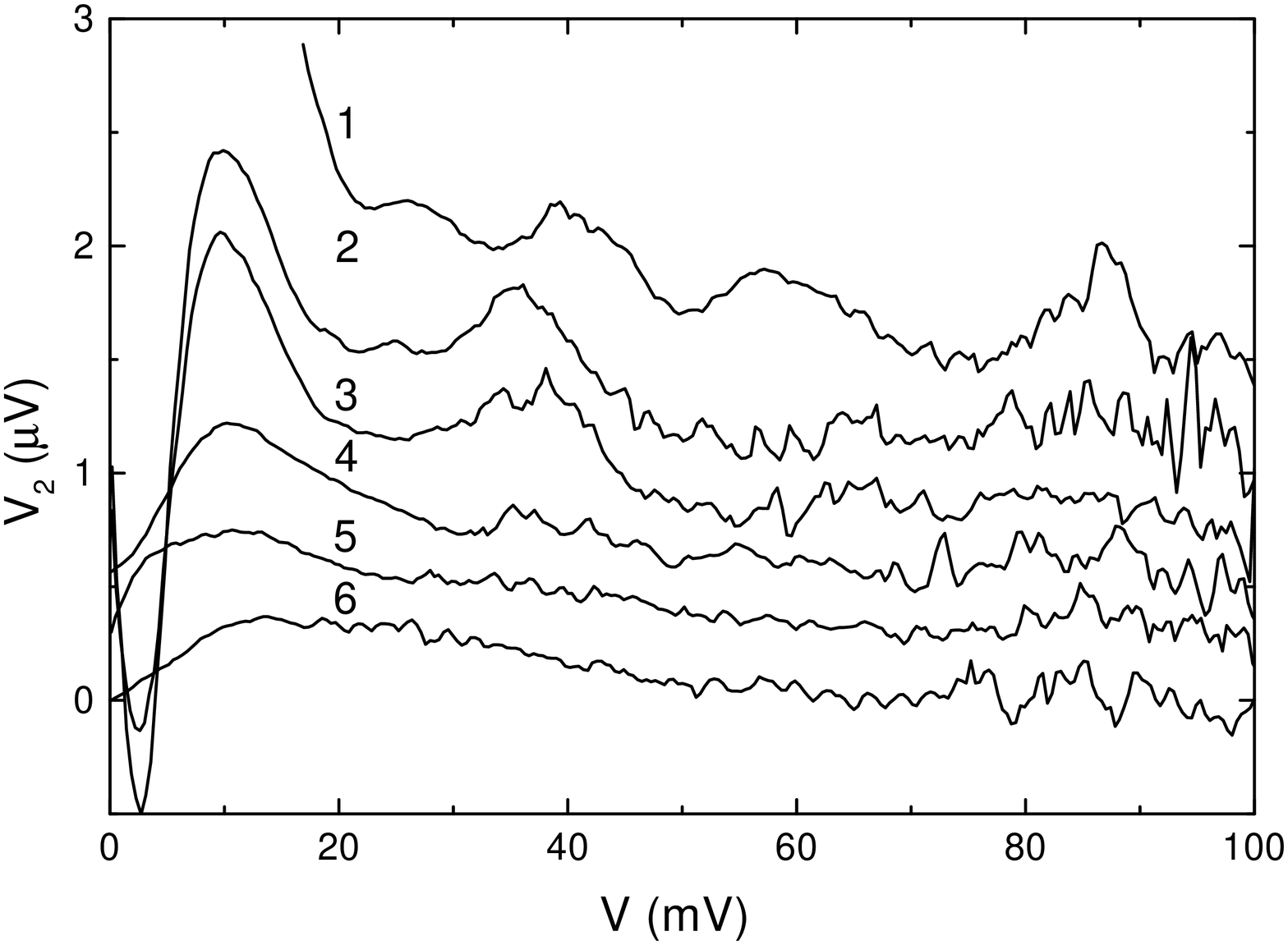}
\end{center}
\caption[] {Second harmonic dependences on field and temperature
averaged for both voltage polarities. The temperature and magnetic
field are: 4.2\,K, 0\,T; 4.2\,K, 4\,T; 10\,K, 4T; 20\,K, 4\,T;
30\,K, 4\,T; 40\,K, 4\,T, for curves 1-6, respectively. The
modulation voltage $V_1$(0) is 2.2\,mV. $\Delta$=2.7\,meV. Curves
are shifted vertically for clarity. } \label{yansf6}
\end{figure}

{\it Conclusion}. We have measured the PC spectra in transition
metal diborides: ZrB$_{2}$, NbB$_{2}$ and TaB$_{2}$. For all
compounds the main phonon peak position was established, the PC
EPI function was recovered, and the EPI parameter $\lambda$ was
determined. The obtained small $\lambda$ values question strongly
the reported bulk superconductivity in these compounds with
remarkable $T_c$. To draw a more weighty conclusion about details
of EPI and the $\lambda$ values in the presented MeB$_2$ family, a
theoretical calculation of $\alpha_{\rm PC}^2 F(\omega)$ with the
mentioned $K$-factor and their comparison with experimental data
is very desirable.

The PC EPI spectra of the superconducting MgB$_2$ differ even
qualitatively from those measured for the mentioned
transition-metals diborides. For MgB$_2$ the reproducible peaks on
$d^{2}V/dI^{2}(V)$ are seen in the superconducting state, which
disappear in the normal state. It proves that they are due to the
energy dependence of the superconducting order parameter. The
superconductivity here is presumably due to the 2D-band EPI, which
induces the $\Delta (E)$ structure in the 3D band. At a small
value of the superconducting gap the phonon structure is weak, and
its intensity begins to increase with growing $\Delta$. The
robustness against the magnetic field also grows notably with
increasing $\Delta$.

In the normal state the intensity of the inelastic spectrum also
correlates with the value of the superconducting gap. The tendency
is observed that the high energy phonon contribution to the PC
spectra becomes more prominent for a larger gap. This is supported
also by our recent data on MgB$_2$ single crystals\cite{mgbphon}.
For a smaller gap, the low-frequency bump appears in the spectra.
It might be due to EPI with some unknown low-frequency
excitation\cite{Lampakis}, but further work is needed to exclude
that it is not caused simply by phonon peaks of the normal-metal
counter electrode.


\end{document}